\documentclass{jetpl}
\usepackage{graphicx}
\twocolumn
\lat
\title{Steps on current-voltage characteristics of a silicon 
quantum dot covered by natural oxide}
\rtitle{Steps on current-voltage characteristics\ldots}
\sodtitle{Steps on current-voltage characteristics of a silicon 
quantum dot covered by natural oxide}
\author{S.\,V.\,Vyshenski$^{+}$\/\thanks{e-mail: 
svysh@pn.sinp.msu.ru},
U.\,Zeitler$^{*}$, and R.\,J.\,Haug$^{*}$}
\rauthor{S.\,V.\,Vyshenski, U.\,Zeitler, and R.\,J.\,Haug}
\sodauthor{Vyshenski, Zeitler, Haug}
\address{$^+$Nuclear Physics Institute, Moscow State University, 
Moscow 119899, Russia\\~\\
$^*$Institut f\"ur Festk\"orperphysik, Universit\"at Hannover,
Appelstr. 2, D-30167 Hannover, Germany}
\dates{2 October 2002}{*}
\abstract{Considering a double-barrier structure formed by a silicon 
quantum dot covered by natural oxide with two metallic terminals, we 
derive simple conditions for a step-like voltage-current curve. Due 
to standard chemical properties, doping phosphorus atoms located in 
a certain domain of the dot form geometrically parallel current 
channels. The height of the current step typically
equals to $N \times 1.2$\,pA, where $N=0,1,2,3\ldots$ is the number
of doping atoms inside the domain, and only negligibly depends on 
the actual position of the dopants. The found conditions are 
feasible in experimentally available structures.}
\begin{document}
\maketitle

The fabrication of $Si$ nanostructures became possible through
recently developed new technologies \cite{Chou,Oda}.
Individual silicon quantum
dots (SQD) reported in \cite{Oda} are spherical
$Si$ particles with diameters $d$ in the range 5\ch12\,nm covered
by a 1\ch2\,nm-thick natural $SiO_2$ film. Metallic current
terminals made from degenerately doped $Si$ are defined lithographically
to touch each individual dot from above and from below.

To ensure metallic electrodes the donor concentration $n$
should be $n\ge n_{Mott}$,
where $n_{Mott}=7.3\times 10^{17}$\,cm$^{-3}$. The critical 
concentration
$n_{Mott}$ is defined by the \emph{Mott criterion} \cite{Mott},
introducing the transition to a metallic type of conductivity
in a semiconductor at:
\begin{equation}
a_B\times (n_{Mott})^{1/3}=0.27 .  \label{Mott}
\end{equation}
where $a_B$ is the Bohr radius of an electron
bound to a donor inside the $Si$ crystal, in the case of
phosphorus-donors $a_B = 3$\,nm \cite{Mott}.

As for the doping of the dot, the situation concerning a
Mott transition in that small dots is much less trivial than the
one described by Eq. (\ref{Mott}).
Let us consider dots with diameters $d=10$\,nm formed from n-doped
$Si$ with $n = n_{Mott}$ as an illustrative example.
Then each dot contains in average one donor. Note that we will consider
degenerately $n^+$-doped electrodes with $n \gg n_{Mott}$ which ensures
metallic conduction up to the borders of the dot.

Real fabrication technology \cite{Oda} provides a wafer with hundreds of
SQDs on it with current leads towards each individual SQD. Dots in average
have the same value of mean dopant concentration $n$, which is
determined by the parent material of bulk silicon the dots are formed
from. However, on the level of each individual
SQD we will always have exactly {\emph integer} number of doping atoms. If,
as in the example above, the average number of dopants
$\overline{N_{tot}}=1$ the actual number  of donors
in the dot can have values $N_{tot} = 0, 1, 2, 3, \ldots$, with
values larger than these very unlikely.

Our objective is to illustrate, that SQDs from the same wafer fall
into several distinct sets of approximately the same conductance. The
typical value of conductance for each set is nearly completely
determined by the number $N$ of donors present in a certain part of a SQD
so that $N$ labels each set of SQDs.

Summarizing the above, we need for a quantization of the
conduction through a dot with $N$ donors the following conditions:

\begin{itemize}
\item Size $d$ of the dot comparable with Bohr radius: $2<d/a_B<5$.
\item Average doping $n$ of the dot $n \le d^{-3}$,
leading to a mean number of dopants  $\overline{N_{tot}} \le 1$,
so that $N_{tot}=0, 1, 2$ are the most probable configurations
of an individual SQD.
\item Doping of the electrodes $n_{el} \gg n_{Mott}$,
so that current leads are perfectly metallic.
\item Dot covered by an oxide layer thick enough to suppress ballistic
transport through the dot.
\end{itemize}

In fact all these condition can be simultaneously satisfied
for SQD fabricated with the method mentioned above \cite{Oda}.

\begin{figure}[htp]
  \centering
  \includegraphics[width=80mm]{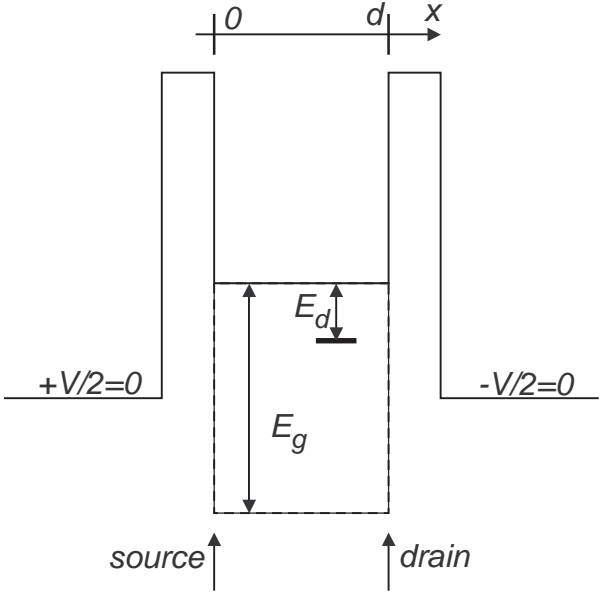}
  \caption{Fig.\ref{fig:fig1}. Potential profile of the dot covered 
by an oxide layer at $V=0$. A donor is marked with a short bar}
  \label{fig:fig1}
\end{figure}

\textbf{Model system.} We use a simple model of a cubic SQD with 
$d > 2 a_B$ (we will use
$d=10$\,nm for estimates), covered with an
oxide layer with thickness $\delta = 2$\,nm, height \cite{refbook}
$B = 3.15$\,eV and contacted with current
terminals from left and from right.
The $x$-axis is oriented from left to right along the current flow,
as shown in Fig.\ref{fig:fig1}. 

\begin{figure}[htp]
  \centering
  \includegraphics[width=80mm]{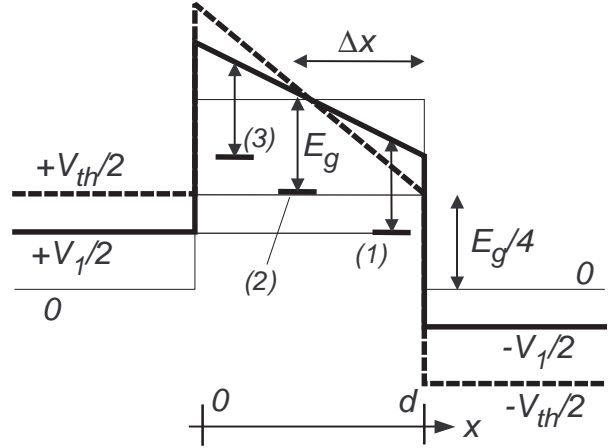}
  \caption{Fig.\ref{fig:fig2}. Potential profile of the dot between 
effective source and drain biased with $V_{eff}=V_1$ 
(thick solid line) and $V_{eff}=V_{th}$ (dashed line)}
  \label{fig:fig2}
\end{figure}

A tunneling current is injected into the dot via the oxide barrier from
the top (source at $x=0$) and leaves the dot at the bottom
(drain at $x=d$). Due to the presence of the oxide barriers
this current is non-ballistic and non-thermal.
We assume that the high potential barriers associated with the
oxide layers are not much affected by the voltage and the tunneling
charges. We concentrate on what happens between these effective source 
and drain (Fig.\ref{fig:fig2}), as in \cite{Schw}.

In the case when the dot can be regarded as an insulating system
it is reasonable to assume that the applied voltage equally
drops over the potential barriers and the dots. For simplicity
we neglect the difference of the dielectric constants of the
oxide barriers and the dot.
In this approximation we can introduce
an effective voltage  $V_{eff}=V (d-2 \delta) / d = 0.6\  V$
describing the part of the total
transport voltage $V$ applied between effective source and drain
which drops across the dot itself.

In this rude approximation we neglect the effect of spatial quantization
upon values on the ionization energy, the
conductivity gap and material parameters of silicon.

\textbf{Dot without donors.} At $V_{eff}=0$ the Fermi level inside the dot
is situated in the middle of the gap, i.e. $E_g/2$ bellow
the conduction band edge ($E_g=1.14$\,eV at 300 K).

As $V_{eff}$ grows, the bottom of the (still empty) conduction band bends
down accordingly. When the conduction band in the dot close to the drain
aligns with the Fermi level of the emitter we expect a drastic
increase in the tunneling current. This threshold $V_{th}$ voltage (Fig.\ref{fig:fig2})
for $V_{eff}$ is given by
$V_{th} = E_g/(2 e)$,
regardless of the number $N_{tot}$ of dopants in the dot
(as long as the dot is not yet metallic, of course).
In the following we therefore limit our
studies to voltages

\begin{equation}
\left\vert V_{eff} \right\vert \le V_{th} = E_g/(2 e) = 0.57 \mbox{\,V}.
\label{threshold}
\end{equation}

In this voltage range we have a
$d$-thick barrier (formed by the dot) with always finite height
between effective source and drain. The intrinsic
concentration of electrons and holes at 300 K is $1.4 \times 
10^{10}$\,cm$^{-3}$. Even at this high temperature the 
probability to have at
least one intrinsic electron in a dot
with size $d=10$\,nm is only $1.4 \times 10^{-8}$.
So we would expect virtually no
current in this mode. This
is confirmed by direct electrical tests \cite{Oda}
of SQD with the required properties.

\textbf{Single-donor channel.} Let us now consider one single
donor in the dot located at $x$ with ionization energy \cite{refbook}
$E_d=0.045$\,eV (for P as a donor).

At zero temperature current is due to resonant tunneling via non-ionized
donor (as in \cite{Larkin} for example). Differential conductance
$g(\varepsilon)$ for the states with energy $\varepsilon$ is

\begin{equation}
g(\varepsilon) = \frac{4 e^2}{\pi \hbar}
\frac{\Gamma_\ell \Gamma_r}{(\varepsilon-\varepsilon_d)^2+
(\Gamma_\ell + \Gamma_r)^2},
\label{g}
\end{equation}
where $\Gamma_{\ell,r}$ is linewidth of $1s$ state of electron bound to the
donor due to coherent mixing with conduction states
to the left (right) of the left (right) tunnel barrier.

Oxide barriers (with height $B=3.15$\,eV
and width $\delta=2$\,nm) give dominant contribution to $\Gamma_{\ell,r}$
compared to contribution of the body of the dot (with
typical height $<E_g/2=0.57$\,eV and width $<d=10$\,nm).
So, we can approximate $\Gamma_{\ell,r}$ with linewidth
$\Gamma$ for the case of an impurity localized at distance $\delta$ inside
rectangular one-dimensional tunnel burrier \cite{Larkin} of height $B$:

\begin{equation}
\Gamma_\ell = \Gamma_r = \Gamma = \frac{2 p_F \kappa}{p_F ^2+\kappa^2}
B \frac{exp(-2\kappa \delta)}{\kappa \delta} = 2.5 \times 10^{-9} 
\mbox{\,eV},
\label{G}
\end{equation}
where $m$ is (true, not effective) electron mass, $\kappa=(2mB)^{1/2} 
/ \hbar$,
and $p_F =(2mE_F)^{1/2}/\hbar=
\left( 3\pi^2n_{el} \right) ^{1/3}$ is Fermi wave number in the contact 
electrodes.
The numerical estimate in (\ref{G}) is given for the  electrodes doped
up to $n_{el} = 10^{21} \mbox{\,cm}^{-3}$  as in \cite{Oda}.

Within approximation (\ref{G}) point $\varepsilon = \varepsilon_d$ 
brings
function $g(\varepsilon)$ given by (\ref{g}) to a sharp maximum
$g(\varepsilon_d) = e^2/\pi \hbar$ of width $\Gamma \ll V_th$.

From Fig.\ref{fig:fig2} it is clear that resonant energy
$\varepsilon_d = E_g/2 - x V_{eff}/(2 e d) - E_d$.
This means that as soon as effective Fermi level $eV_{eff}/2$ reaches
a certain threshold $eV_1/2$, tunnel current $J$ flowing through the 
structure acquires a step-like increase of

\begin{equation}
J_1 = \frac{1}{e}\int_{-eV/2}^{eV/2} g(\varepsilon) d \varepsilon =
\frac{g(\varepsilon_d) 2 \pi \Gamma}{e} = \frac{2 e \Gamma}{\hbar} = 
1.2 \mbox{\,pA},
\label{G1}
\end{equation}

If  the impurity is located near the drain,
i.e. $d-a_B < x < d$ (as donor 1 in Fig.\ref{fig:fig2}), then threshold $V_1$
for the effective voltage $V_{eff}$ is given by
\begin{equation}
V_1=E_g/(2 e)-E_d/e= 0.525 \mbox{\,V}.
\label{V1}
\end{equation}

In contrast, for an impurity located at distances $\Delta x>2 d 
E_d/E_g $
from the drain (i.e. further away than the threshold
case of donor 2 in Fig.\ref{fig:fig2}),
no additional current channel via a single impurity can be
opened at low enough voltages defined in (\ref{threshold})
where virtually no background current is present.
In the present case this value $\Delta x =
0.8$\,nm, which returns us to the above criterion: only impurities 
located in
the immediate vicinity (defined within the accuracy $a_B$) of the 
drain
contribute to the single-impurity channel.

This shows that in first approximation the conductance of this 
channel does not
depend on $x$. As shown above, a single-impurity channel already
only selects impurities located within a very narrow range of $x$
close to the drain.

\textbf{Two-, three-, multi-donor channel.} The above consideration 
shows, that due to the
bend of the bottom of conduction
band following the transport voltage, there is no chance to notice 
current
flowing through a sequential chain of impurities (such as donors 1 
and 3 in
Fig.\ref{fig:fig2}), connecting source and
drain. The contribution of such a chain will be totally masked by
the current flowing directly
via the conduction band. The only way for multiple impurities to 
manifest
themselves in quantized conductance is to form multiple
\emph{geometrically parallel} singe-impurity channels situated close 
enough to the drain as considered
above.

\begin{figure}[htp]
  \centering
  \includegraphics[width=80mm]{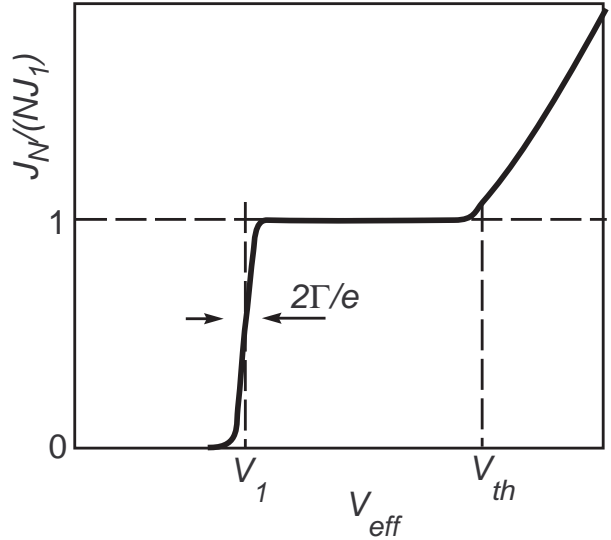}
  \caption{Fig.\ref{fig:fig3}. Current-voltage characteristics of a 
model system (not to scale)}
  \label{fig:fig3}
\end{figure}

Therefore, if $N>1$ impurities fall into the thin layer
near the drain  to approximately the same $x$ coordinate
as that of donor 1 in Fig.\ref{fig:fig2} (within the Bohr radius), we will see a
switching-on of an $N$-fold channel with current
\begin{equation}
J_N=N J_1 = N 2 e \Gamma / \hbar = N \times 1.2 \mbox{\,pA}
\label{GN}
\end{equation}
at the same
threshold voltage $V_{eff}= V_1 = 0.525$ V as for a single-donor 
channel (Fig.\ref{fig:fig3}).

All the above considerations are only valid as long as the dot itself
can be regarded as an insulating system.
As the number of donors in a SQD grows, the dot becomes a metallic
particle, and the
conduction band edge in the dot aligns with the Fermi level of the
electrodes.
In a very simple estimate we define this transition to a metal when
the total volume of $N_{tot}$ donors with an individual volume
of $4\pi/3 \times a_B^3$ exceeds the volume of the dot.
This is an exaggerated version of the Mott criterion (\ref{Mott}) 
which holds
not only in bulk, but in a small structure, too.
For the analyzed example from above this gives
$N_{tot}=8$ as a limiting value. The
practically interesting set $0,1,2,3,\ldots$ for both $N_{tot}$ 
and $N$
considered above is still far bellow this limit.

Quite a number of other mechanisms of electron transport might take 
place in this system. Surprisingly,
even taking into account such other mechanisms \cite{Kupriyanov}
does not change much the main idea of the present paper.

In small dots with diameter $d < 2 a_B = 6$\,nm the domain with $N$ 
active dopants extends to the whole dot, and thus $N=N_{tot}$. In 
large dots with diameter $d > 2 a_B = 6$\,nm the domain with active 
dopants is less than the dot itself and is localized near the drain. 
Hence the position of the domain, number $N$ and value $J_N$ all can 
be different for the current flowing in different directions. Really, 
when sign of applied voltage changes, the source and the drain 
change places.

In a certain sense the discrete increase of dot's conductivity which 
follow the increase of the dopants number could be regarded as a 
\emph{mesoscopic analog of the Mott transition} between insulating 
and conducting states of the system.

Useful discussions with I. Devyatov, M. Kupriyanov, and  S. Oda 
are gratefully acknowledged.

\vfill\eject


\begin{thebibliography}{99}

\bibitem{Chou} L.\,J. Guo, E. Leobandung, L. Zhuang et al.,
J. Vac. Sci. Technol. B \textbf{15}, 2840 (1997).

\bibitem{Oda}  A. Dutta, M. Kimura, Y. Honda et al.,
Jpn. J. Appl. Phys. {\textbf 35}, 4038 (1997). S. Oda,
K. Nishiguchi, Journal de Physique IV,  \textbf{11} (Pr.3), 
1065 (2001).

\bibitem{refbook} O. Madelung (ed), \emph{Physics of Group IV 
Elements and III-V Compounds}
(Springer-Verlag, Berlin, 1982), Subvolume III/17a of 
Landolt-B\"ornstein New Series.

\bibitem{Mott} N.\,F. Mott, \emph{Metall-Insulator Transitions}
(Taylor \& Francis, London, 1990, 2-nd edition), Chap. 5.

\bibitem{Schw} J. Schwinger, \emph{Particles, sources, and fields} 
(Addison-Wesley,
Reading, 1970).

\bibitem{Larkin} A.\,I. Larkin and K.\,A. Matveev, Zh. Exp. Theor. 
Phys. \textbf{93}, 1030 (1987).

\bibitem{Kupriyanov} I.\,A. Devyatov and M.\,Yu. Kupriyanov, 
JETP \textbf{77}, 874 (1993).

\end{thebibliography}
\end{document}